\documentclass{JHEP3}

\usepackage{graphicx}
\usepackage{amssymb}
\usepackage[mathscr]{eucal}
\usepackage{amsmath}
\usepackage{cite}
\usepackage{afterpage}

\newcommand{\bc}{\begin{center}}
\newcommand{\ec}{\end{center}}
\newcommand{\bi}{\begin{itemize}}
\newcommand{\ei}{\end{itemize}}
\newcommand{\ii}{\item}

\usepackage{amssymb}
\usepackage{float,graphicx}

\newcommand{\bfig}{\begin{figure}}
\newcommand{\efig}{\end{figure}}

\newcommand{\Ccal}{\mathcal{C}}
\newcommand{\be}{\begin{eqnarray}}
\newcommand{\nn}{\nonumber\\}
\newcommand{\ee}{\end{eqnarray}}
\newcommand{\bq}{\begin{equation}}
\newcommand{\eq}{\end{equation}}
\newcommand{\ba}{\begin{array}{ccc}}
\newcommand{\ea} {\end{array}}
\newcommand{\bv}{\begin{array}{c}}
\newcommand{\ev} {\end{array}}

\newcommand{\N}{{\cal{N}}}

\newcommand{\gsimm}{\raise.3ex\hbox{$>$\kern-.75em\lower1ex\hbox{$\sim$}}}
\newcommand{\lsimm}{\raise.3ex\hbox{$<$\kern-.75em\lower1ex\hbox{$\sim$}}}

\newcommand{\la}{\label}
\newcommand{\f}{\frac}

\title{Gauge/Cosmology Brane-to-Brane Duality}

\author{Philippe Brax and Robi Peschanski\\
  Institut de Physique Th\'eorique, CEA, IPhT, CNRS, URA 2306,
  F-91191Gif/Yvette Cedex,\\ E-mail:
  \email{philippe.brax@cea.fr, robi.peschanski@cea.fr}}

\date{today}

\abstract{We introduce a duality relation  between {{two {\it distinct} branes}}, a {\it cosmological} brane with macroscopic matter and a {\it holographic} brane with microscopic gauge fields. Using  brane-world cosmology with a single brane in a 5-dimensional $AdS_5$ background, we find an explicit time-dependent holographic correspondence between the bulk metric surrounding the cosmological brane and the $\N=4$ gauge field theory living on the  boundary of the $Z_2$-symmetric  mirror bulk,  identified with  the holographic brane. We then relate  the cosmic acceleration on the cosmological brane  to the conformal anomaly of the gauge theory on the holographic brane. This  leads to a dual microscopic interpretation of the number of e-foldings of the cosmological eras on the cosmological brane.}

\preprint{...}
\begin{document}

\section{Introduction}
The Gauge/Gravity duality, relating a gauge field theory at strong coupling on a 4 dimensional brane to a weakly-coupled gravitational theory in the 5-dimensional bulk with the brane at its boundary, has found a large number of applications. In the guise of the AdS/CFT correspondence between the $\N=4$ Super Yang-Mills (SYM) {{conformal field}}  theory and metrics and matter on an Anti de Sitter 5-dimensional space \cite{adscft}, it has led to precise Gauge/Gravity duality relations, including time-dependent configurations \cite{JaniR}.

Most practical applications of the AdS/CFT correspondence concern in effect its ability to connect, on the one hand observables of the microscopic gauge theory whose strong coupling regime prevents  direct calculations, with, on the other hand,  practical evaluations in  the (super) gravity approximation on a background with weak curvature.

In the present paper, our goal is the converse, namely using the AdS/CFT correspondence to obtain  some new insight on open cosmological problems in 4-dimensional gravity. In practice, we will  consider that the bulk gravity description gives information about two dual aspects of the same physical system: on the one hand, physical cosmological features in 4 dimensions and  gauge theoretic properties on the other hand. Even if the strong coupling limit of a gauge field theory is  difficult to describe, some general aspects can be unraveled, in particular with the help of the AdS/CFT correspondence, therefore allowing one to shed new light on 4d cosmology. On more general grounds, investigating the  possible  realisation of a microscopic dual to large scale  cosmological properties is a stimulating challenge {\it per se}.

Our construction will make use of  cosmological brane-world models \cite{Bine,brax} where the conventional 4-dimensional cosmology is realised on a single {\it cosmological} brane surrounded by an $AdS_5$ background metric. We will look for a holographic Gauge/Gravity duality between the {\it bulk} 5-dimensional metric of the brane-world model and the gauge theory on a {\it holographic} brane, which is {\it distinct} from the {\it cosmological} brane, while both embedded in the same 5-dimensional bulk. This will be made possible upon identifying within this brane-world scenario a known class of time-dependent holographic duals \cite{Kajan}. In this geometrical set-up, as the holographic  brane is distinct from the cosmological brane,  we  will establish a brane-to-brane duality mediated by the 5-dimensional bulk. More precisely, we focus on brane world models where the cosmological  brane is located at $z=1$ in Feffermann-Graham~\cite{Feff} coordinates and the physical 5d bulk extends to $z=\infty$. The brane world system is such that cosmological matter is embedded on the $z=1$ physical brane. The Israel junction conditions relate the matter content on the physical brane to its extrinsic curvature. We also assume that the physical brane is the fixed point of a $Z_2$ symmetry exchanging $z$ and $1/z$. The mirror image of the bulk is associated to  the unit interval $z\in[0,1]$ whose boundary at $z=0$ is identified with the holographic brane associated with the mirror bulk viewed as the time dependent holographic dual. The duality which will be spelt out in the rest of the paper connects a classical brane at $z=1$ whose matter content is related to the  (mirror) bulk metric via the Israel conditions and the holographic brane whose matter content is dual to the (mirror) bulk metric thanks to the holographic renormalisation group\cite{Sken}. In a sense, this realises a classical-quantum duality which will be more transparent when we relate a Heisenberg-like uncertainty relation to the classical number of efoldings on the physical brane.

Our  goal is then to try and provide  new insight on  standard issues in cosmology, using the Gauge/Cosmology duality obtained thanks to holographic properties of brane-world models.  We  focus mainly on the acceleration of the universe. Indeed these  holographic tools may lead to  a new approach to   the cosmological constant problem. Our model differs significantly from the holographic dark energy setting whereby a modification of the Friedmann equation which is sensitive to the IR cut-off of the theory is introduced\cite{holo}. Holography has also been used to interpret the effective brane equations of motions as dual to the bulk metric\cite{dual}. Here the dynamics on the physical brane are dual to a gauge field theory on a separate brane via the (mirror) 5d bulk.

In section \ref{AdS}, we recall some basic elements of  $AdS_5$~brane-world cosmological models \cite{Bine,brax}, where the low energy cosmology can be reproduced on a {\it cosmological} brane. In the following section \ref{HolO}, we introduce the {\it holographic} brane, as the boundary of the $Z_2$ symmetric $AdS_5$ space-time surrounding the cosmological brane, in the simpler case of the absence of dark radiation, and analyse its duality properties. The main section \ref{GCD} gives the explicit time-dependent $AdS_5/CFT$ correspondence including  dark radiation which provides the full Gauge/Cosmology duality framework between the $\N=4$ SYM theory on the (curved) {\it holographic} brane and the classical gravitational regime  on the {\it cosmological} brane. We emphasize that  a physical interpretation of the Gauge/Cosmology duality is  a microscopic derivation of  the number of e-foldings of the cosmological eras. In the final section \ref{Fine} we summarise our results.
\section{The $AdS_5$ brane cosmology}
\la{AdS}
\bfig[ht]
\bc
\includegraphics[width=7cm]{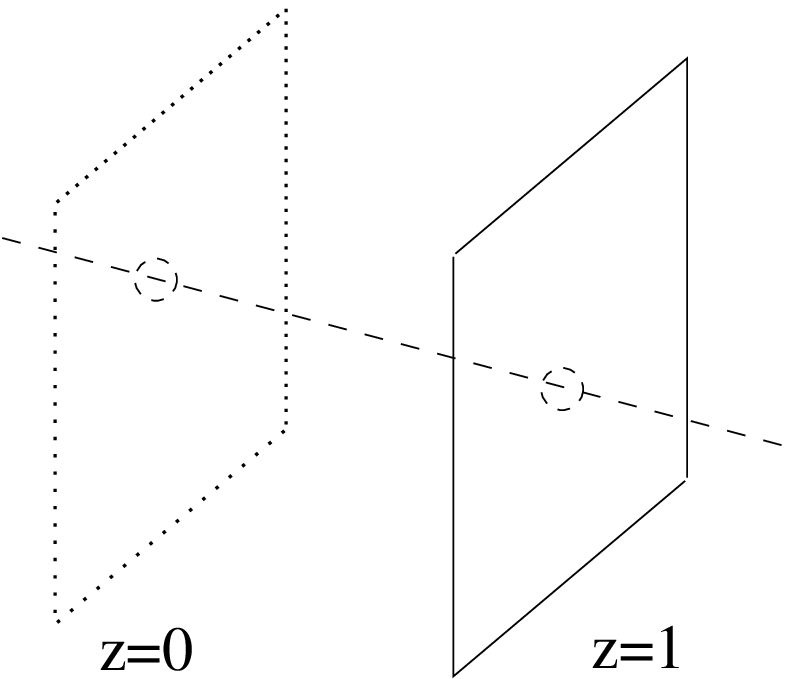}
\ec
\caption{{\it The cosmological and holographic branes.} At $z=1:$ the cosmological brane;  and at $z=0:$ the holographic brane. The  5-dimensional $AdS_5$ bulk, with its symmetry $z\to  1/z,$ is depicted using the $z$-variable of the Fefferman-Graham metric as the axis orthogonal to the branes. The physical bulk corresponds to $z\ge 1$ while the mirror bulk is $z\in [0,1]$. The boundary $z=0$ is the holographic brane and the gauge/cosmology duality operates via the mirror bulk from $z=0$ to $z=1$ and vice versa.}
\la{1}
\efig

\subsection{The bulk-brane cosmological geometry}
The presence of a boundary brane carrying matter embedded in a $AdS_5$ bulk leads to a cosmological dynamics with a non-static bulk \cite{Bine} (see also \cite{Kiri,Cvet}. The time evolution of the bulk metric is related to the time variation of the energy momentum tensor on the brane. Here, we focus on an interval $y\ge 0$ isometric to $AdS_5$ with a boundary brane at $y=0$  ($z=1$ in Fig.\ref {1}, using Fefferman-Graham coordinates). Before including matter on the boundary brane, the bulk metric and therefore  the fifth dimension geometry is that of the vacuum $AdS_5$ specified  by  a metric
\be
ds_5^2=dy^2-e^{2y/l}\ (dt^2-dx^2)\
\la{Vac}
,\ee
where the 3d metric is $dx^2\equiv \sum_1^3 dx_i^2$ and $l$ is the $AdS$ radius.

The metric in the bulk equipped with the boundary brane must satisfy the Einstein equations while the metric on the boundary satisfies the Israel junction conditions. The Einstein equations read:
\begin{equation}
R_{AB}-\frac{1}{2} R{{\ }} g_{AB}= \kappa_5^2\ T_{AB}
\end{equation}
where $A=\{1\dots 5\}$ label  the 5d coordinates and the energy momentum tensor is purely diagonal
\begin{equation}
T^A_B={\rm diag} (-\rho_B,\rho_B,\rho_B,\rho_B,\rho_B)\ ,
\end{equation}
where $\rho_B$ is the energy density in the bulk. The Israel conditions relate the extrinsic curvature tensor of the boundary brane at $y=0$ to the brane energy-momentum tensor
\begin{equation}
K_{AB}=-\frac{\kappa_5^2}{2} (T_{AB}^m-\frac{1}{3}T^mh_{AB})
\end{equation}
where $K_{AB}= h_A^Ch_B^D \nabla_C n_D$ is the intrinsic curvature, $h_{AB}= g_{AB}-n_An_B$ the induced metric on the brane, $\nabla_A$ the covariant derivative for the 5d metric and $T^m_{AB}$ the matter energy momentum tensor on the brane. The vector $n_A$ is the normal vector to the brane at $y=0$ and satisfies $n_An^A=1$. Notice that $K_{AB}n^A=0,\ h_{AB} n^B=0,\ T^m_{AB}n^B=0,$ implying that the Israel conditions are effective equations along the brane in 4d only.

One then assumes that the boundary brane is the fixed point of a $Z_2$ symmetry allowing one to identify both $y$ and $-y$. In this context, and in normal coordinates we choose  the brane energy momentum tensor along the 4d directions to be
\be
T^{m,a}_b={\rm diag} (-\rho_b,p_b,p_b,p_b)\
\la{Bran}
\ee
where $\rho_b$ is the brane energy density and $p_b$ the brane pressure.

In the single  brane cosmological model \cite{Bine}, one uses the embedding of a single 3-brane in five dimensions when both the brane and the bulk have a cosmological constant. One assumes that the bulk
cosmological constant and the brane tension are related. This relation is akin to the one in the Randall-Sundrum model \cite{RS} and guarantees the existence of static bulk solutions in the absence of brane matter. This choice amounts to
\begin{equation}
\kappa_5^2\rho_B=-\frac{6}{l^2}
\end{equation}
where $\kappa_5^2= 1/M_5^3$ is the 5d gravitational constant and $l$ is the bulk AdS radius or inverse curvature, together with the identification
\begin{equation}
\kappa_5^2 \rho_\Lambda=\frac{6}{l}
\end{equation}
where $\rho_b=\rho_\Lambda$ when no matter is present.
In this case, this implies  that the brane metric is the Minkowski metric.
 The 4d gravitational constant is
\begin{equation}
8\pi G_N\equiv \kappa_4^2= \frac{\kappa_5^2}{l}
\end{equation}
relating 4d and 5d gravitational interactions. Phenomenologically, the constraint that Big Bang Nucleosynthesis (BBN) should not be altered yields a conservative bound $\rho_\Lambda \ge (1 {\rm MeV})^4$. When we consider low energy phenomena, we will always focus on scales much lower than $\rho_{\Lambda}$.

\subsection{The cosmological brane dynamics}
In normal coordinates, the normal to the brane $n_A$ points along the $y$ direction only and the bulk metric reads
\begin{equation}
ds^2= dy^2 -n^2(t,y) dt^2 + a^2(t,y) dx^2
\la{Bulk}\end{equation}
where $n$ and $a$ depend both on $t$ and $y$. The Einstein equations relate $n$ to $a$ and $a_0$. The scale factor on the boundary brane is:
\begin{equation}
n=\frac{\dot a}{\dot a_0}
\la{Eqn}
\end{equation}
with a brane metric at $y=0$ reading
\begin{equation}
ds^2_b= -\ dt^2 + a_0^2 dx^2
\end{equation}
The dynamics of the brane  are determined by the Israel conditions.
The Israel junction conditions  imply that the scale factor on the boundary brane satisfies a Friedmann equation where $H=\frac{\dot a_0}{a_0}$
is the Hubble factor such that
\begin{equation}
H^2l^2=-1 + \frac{\rho_b^2}{\rho_\Lambda^2}+ \frac{\Ccal}{a_0^4}\ ,
\end{equation}
where the {\it a priori}  free parameter $\Ccal$ corresponds to a  ``dark radiation'' contribution \cite{Bine}. The brane matter density is identified with
\begin{equation}
\rho_b= \rho_\Lambda + \rho
\end{equation}
where  $\rho$ is the cosmological matter density and $p_b= -\rho_{\Lambda} + w \rho$ where $w$ is the brane matter equation of state.
At low energy when $Hl\ll 1$, the Friedmann equations reduces to
\begin{equation}
H^2l^2= \frac{2\rho}{\rho_{\Lambda}} + \frac{\Ccal}{a_0^4}\ ,
\end{equation}
which is valid as long as $\rho\ll \rho_{\Lambda}$. Moreover, the dark radiation term $\frac{\Ccal}{a_0^4}$ should not exceed 10\% of the matter content during BBN.

For further use, we will switch to the Fefferman-Graham coordinates \cite{Feff} which are more suitable for holographic properties \cite{Sken}. They are easily obtained by the change of variable $z=e^{y/l}.$ Then the metric \eqref{Bulk} becomes:
\begin{equation}
ds^2= {l^2}\ \frac {dz^2}{z^2} -n^2(z,t)dt^2 + a^2(z,t)dx^2
\la{Eqs}
\end{equation}
and  the metric {{\cite{Bine}}}  takes the form
\be
a^2&=&\f {a_0^2}{z^2}\ \left\{ -z^2 {H^2l^2}/{2} - \frac{1}{2} (z^4 -{1})\sqrt{1-{\Ccal}/{a_0^4}+H^2l^2} +\frac{1}{2}(z^4 +{1}) (1+{H^2l^2}/{2})\right\}\nn
n^2&=&\left[\f{\dot a (z,t)}{\dot a_0(t)}\right]^2\ .
\la{Eqa}
\ee
 Notice that $z$ varies between $1$ and $\infty$ here.
When $H=0$ and $\Ccal=0$ , this metric corresponds to the warped and static $AdS_5$ metric \eqref{Vac} with $a= a_0/z$.

\section{The Holographic Correspondence}
\la{HolO}

\subsection{The holographic brane}

So far we have studied the branch $z\ge 1$ corresponding to $y\ge 0$. Let us now study what happens when $y\le 0$. The $Z_2$ symmetry implies that the solution for $y\le 0$ is obtained by performing $y\to -y$ or equivalently $z\to z^{-1}$ . Hence, in the range  $0\le z\le 1,$ one obtains in Fefferman-Graham coordinates
\be
a^2&=&\f {a_0^2}{z^2}\ \left\{ -z^2 {H^2l^2}/{2} + \frac{1}{2} (z^4 -{1})\sqrt{1-{\Ccal}/{a_0^4}+H^2l^2} +\frac{1}{2}(z^4 +{1}) (1+{H^2l^2}/{2})\right\}\nn
n^2&=&\left[\f{\dot a (z,t)}{\dot a_0(t)}\right]^2\ .
\la{Eqb}
\ee
Comparing  \eqref{Eqb} with  \eqref{Eqa}, the  $Z_2$ symmetry induces a change of sign in front of the second term of $a^2.$

It is well-known that the boundary of the bulk at $y\to -\infty,\ (i.e.~ z\to 0),$ in the cases where a Gauge/Gravity duality is valid,  is the {\it holographic brane} where the gauge theory dual is living. As we shall see further on, this is also exactly realised in our case.

As a first step, let us consider the analytic continuation towards $z=0$ of   the metric \eqref{Eqs} in the simplified  case of $\Ccal=0.$
We get
\begin{equation}
a=\frac{a_0}{2z}\left\vert 1-\sqrt{1+H^2l^2}+z^2(1+\sqrt{1+H^2l^2})\right\vert
\end{equation}
leading to
\begin{equation}
n=\frac{1}{2z\sqrt{1+H^2l^2}}\left\vert 1-\sqrt{1+H^2l^2}+H^2l^2 +\dot Hl^2 -z^2(1+\sqrt{1+H^2l^2}+H^2l^2+\dot H l^2)\right\vert
\end{equation}
Notice that $z\le 1$ here\footnote{Note that the metric is such that $a$ vanishes for $z_i\approx \frac{Hl}{2}\ll 1 $. This singularity will disappear for $\Ccal\ne 0$.}.  In this patch, it is convenient to rewrite the metric
\begin{equation}
ds^2= \frac{l^2}{z^2} (dz^2 + g_{\mu\nu}(x,z) \frac{dx^\mu dx^\nu}{l^2})
\end{equation}
where
\begin{equation}
g_{\mu\nu}(x,z)\ dx^\mu dx^\nu= -n^2 {dt^2} + a^2 {dx^2}
\end{equation}
The expansion of the boundary metric
\begin{equation}
g_{\mu\nu}(x,z)= \sum_{n=0}^\infty g_{\mu\nu}^{(n)} (x) z^{2n}
\end{equation}
is crucial for the holographic set-up \cite{Sken}. Indeed, this allows one to identify the metric on the holographic brane at $z=0$ as
\be
g_{\mu\nu}(x)\equiv g_{\mu\nu}^{(0)}\la{Holom}
\ee
and the energy momentum tensor of the dual gauge theory on the brane as\footnote{The formula \eqref{HoloT} simplifies for a Minkowski boundary, since in that case one can show that $g^{(2)}_{\mu\nu}\equiv 0.$ The full formula is needed for a non flat boundary metric like in our case.}
\begin{equation}
T_{\mu\nu}^H= \frac{2}{l\kappa_5^2} \left\{ g_{\mu\nu}^{(4)} -\frac{1}{8} g^{(0)}_{\mu\nu} \left[ ({\rm Tr } (g^{(2)}))^2 - ({\rm Tr } g^{(2)})^2\right]-\frac{1}{2}
g^{(2)}_{\mu\rho} g^{(0)\rho\sigma}g^{(2)}_{\sigma\nu}+ \frac{1}{4} ({\rm Tr} g^{(2)} ) g^{(2)}_{\mu\nu}\right\}
\la{HoloT}
\end{equation}
where, by definition, $ {\rm Tr }A= g^{(0) \mu\nu}A_{\nu\rho}\ g^{(0) \rho}_\mu$.

\subsection{ Low energy properties}
Let us focus on the low energy properties $Hl\ll 1$.
Using the boundary  metric,  we find that the brane metric in the low energy $Hl\ll 1$ limit corresponds to
\begin{equation}
g_{00}^{(0)}=-\frac{1}{16} (H^2l^2 +2\dot H l^2)^2,\ \ g_{ii}^{(0)}\equiv a^2_H = \frac{a_0^2}{16} H^4l^4\ .
\la{Bmetric}
\end{equation}
 Notice that, at low energy,   we have on the cosmological brane at $z=1$
\begin{equation}                                                         \dot H= -\frac{3}{2} (1+ w)H^2
\la{DotH}\end{equation}
as a function of the physical equation of state.
Similarly, the holographic energy momentum tensor is the one of a perfect fluid
\begin{equation}
T_{\mu}^{\nu}={\rm diag}(-\rho_H,p_H,p_H,p_H)
\end{equation}
where the indices are being raised using the boundary metric $g^{(0)}_{\mu\nu}.$ The pressure and  the energy density are thus identified with
\begin{equation}
\rho_H=\frac{1}{\kappa_5^2 l}\times\frac{192}{H^4l^4}
\end{equation}
and
\begin{equation}
p_H=-\frac{1}{\kappa_5^2 l}\times\frac{64}{H^4l^4}\times \frac{3H^2+ 2\dot H }{H^2 +2\dot H }\ .
\end{equation}
On the holographic brane, in terms of the equation of state, we find
\begin{equation}
w_{eff}\equiv \frac{p_H}{\rho_H}= -\frac{w}{2+3w}
\la{Weff}
\end{equation}
where
$
w\equiv {p}/{\rho}
$ is the physical equation of state on the cosmological brane.

It is instructive to compare the metric behaviour on the  {\it cosmological} (at $z=1$) and {\it holographic} (at $z=0$) branes shown in  Fig. \ref{1}. First of all, notice that a dust-like behaviour on the physical brane $w=0$ leads to a dust-like behaviour $w_{\rm eff}=0$ on the holographic brane. Moreover, a cosmological constant $w=-1$ is also mapped to a cosmological constant $w_{\rm eff}=-1$. These two properties are important in view of the fact that two physical eras in the history of the universe are described by fluids with these equations of state, $i.e.$ the matter and dark energy dominated eras. Finally, a very important relation which will reappear later in its full generality relates $w=-1/3$ and $w_{\rm eff}=1/3$. This implies that a {\it conformal} fluid on the holographic brane ($w_{\rm eff}=1/3,$ radiation-like) is dual to a steady evolving cosmology  ($w=-1/3,$  no acceleration). Indeed the Raychaudhuri equation at low energy reads
\begin{equation}
\frac{\ddot a_0}{a_0}=-(1+3w)\frac{H^2}{2}
\end{equation}
leading to $\ddot a_0=0$. We will see that the relation  between conformality (or the absence thereof) and acceleration can be extended to the regime where dark radiation is present.

As well-known, the scale factor behaves like
\bq
a_0 \sim t^{\frac 2{3(1+w)}}\ .
\eq
As a result  from (\ref{Bmetric}), the proper-time $\tau$ on the holographic brane is
\begin{equation}
\tau_{max}-\tau= \frac{\vert 2 +3 w\vert}{9(1+w)^2} \frac{1}{t}
\la{tau}
\end{equation}
showing that the proper time on the holographic brane is bounded from above by some value $\tau_{max}$ when $t\to \infty.$ Moreover we find
 from \eqref{Bmetric}, on the holographic brane and in terms of the equation of state,
\begin{equation}
a_H (t) \sim t ^{-\f {2(2+3w)}{3(1+w)}}.
\la{aH}
\end{equation}
The behaviour of the holographic scale factor depends crucially on the equation of state. When $w\le -2/3$, the holographic brane is in an expansion phase while it is in a contracting phase when $w\ge -2/3$.
Note that the passage from a decelerated to an accelerated phase  of the universe on the physical brane is associated to a bounce on the holographic brane as the holographic brane goes from a contracting phase to an expanding phase.

So far, we have  made explicit a  correspondence between the holographic and the cosmological branes at low energy, at least in the case with no dark radiation. We will extend this result to the more general case with  dark radiation. In fact this comes from a more general and consistent holographic $AdS/CFT$ correspondence.

\subsection{Time-dependent AdS/CFT}

We will see that the bulk metric \eqref{Eqa}, analytically continued by
the substitution $z\to 1/z$ in the vicinity of $z=0$, corresponds exactly
to the time-dependent gravity dual of an holographic ${\cal N}=4$  Super
Yang-Mills (SYM) gauge theory with a large number $N_c$ of colours. The  most general solution dual to a gauge
theory with $N_c$ colours and a $O(3)$ symmetry corresponds to a regime
 with spherically expanding matter \cite{Kajan}.
The gravity metric corresponding to this  ${\cal N}=4$  Super Yang-Mills (SYM) gauge theory with spherically expanding matter has been obtained
in its full generality \cite{Kajan} and reads:
\be
ds^2=\frac{l^2}{z^2}\left[dz^2-\frac{dt_K^2}{l^2}\f{h^2r^2}{b(t_K,z)}\left(1+A_2z^2+A_4z^4\right)^2+\frac{dx^2}{l^2} b(t_K,z)\right]\ ,
\la{K}
\ee
where, for further use, we denote by $t_K$ the time variable seen from the holographic brane, $r(t_K)$ the spherical expansion coordinate on the brane and $h(t_K)$ a lapse function whose sign is not determined yet.
 The metric depends on
\be
b(t_K,z)=r^2\left\{\left(1-\f{l^2r'^2}{4h^2r^2}z^2\right)^2+\f{z^4}{4r^4z_0^4}\right\}^2
\ ,
\la{b}
\ee
and
\be
A_2=l^2\left(\f{h'r'}{2h^3r}-\f{r''}{2h^2r}\right)\quad\quad
A_4=l^4\left(\f{r'^2r''}{8h^4r^3}-\f{r'^4}{16h^4r^4}-\f{h'r'^3}{8h^5r^3}-\f 1{4r^4z_0^4}\right)\ \ .
\la{A}
\ee
In all these equations, we have defined  $f'={df}/{dt_K}$.

The metric \eqref{K} is a solution of the bulk  Einstein equation. Since the holographic renormalisation \cite{Sken} holds for this genuine duality realisation, the boundary energy momentum tensor is explicitly given by the formula \eqref{HoloT} which has been expounded in the previous section. The gauge-gravity correspondence then leads to the identification\cite{noj}
\begin{equation}
\frac{l^3}{\kappa_5^2}= \frac{N_c^2}{4\pi^2}\ .
\end{equation}
As usual, the  {{(quantum)}}Gauge/(classical)Gravity correspondence holds in  the large $N_c$ limit corresponding to  the bulk radius of curvature being much larger than the 5d Planck scale.

\section{ Gauge/Cosmology Duality}
\la{GCD}
\subsection{Low energy approximation}
As mentioned\footnote{Ref. \cite{Kajan} quotes that the bulk metric is the {{same solution of the Einstein equations as that of}} \cite{Bine,Kiri,Cvet} using different coordinates. In this section, we derive the explicit mapping between both metrics.} in Ref.\cite{Kajan}, there exists a mapping relating the metric \eqref{A} with the one \eqref{Eqa} obtained by analytic continuation from the brane cosmology of \cite{Bine}.

Let us first discuss this mapping in the absence of dark radiation $\Ccal=0$ and at low energy $Hl\ll1$. In the following section, we will extend the correspondence to the general case.

Following \eqref{Bmetric}, the metric reads
\be
ds^2=\frac{l^2}{z^2}\left[dz^2-\frac{dt_B^2}{l^2}\left(\f{H^2l^2/2+\dot{H}l^2}2-z^2\right)^2+\frac{dx^2}{l^2} a_0^2(t_B)\left(\f{H^2l^2}4-z^2\right)^2\right]\ ,
\la{B}
\ee
where, for the sake of convenience we specify the time coordinate as seen from the  physical cosmological brane as  $dt_B$ ($\equiv {dt}$)  and $\dot f \equiv df/dt_B$. We shall identify this metric with the gauge-gravity dual \eqref{K} with $z_0=\infty$. Indeed,
from the spatial metric element $dx^2$, one  gets
\be
r=\f{H^2l^2}4\ a_0(t_B)\quad\quad h=\epsilon_K\f{r'}{Ha_0},
\la{Id}
\ee
where the last identity comes from the choice of the  metric
 without dark  radiation term ($\Ccal=0$). We have introduced $\epsilon_K=\pm 1$ which is determined in such a way that ${dt_K}/{dt_B}>0$, i.e. the flow of time is not reversed on both the holographic and the physical branes.

Indeed,
considering now the first term in $dt_K^2$ from \eqref{K}, one can write
\be
hdt_K = \epsilon_K \f{\dot{r}dt_B}{Ha_0}=\frac{\epsilon_K}{2}\left(\frac{H^2l^2}{2}+\dot{H}l^2\right)dt_B\ ,
\la{first}
\ee
where one made use of \eqref{Id}. This gives the identification
 for the first term in $dt_B$, by comparing \eqref{B} and \eqref{K}, using \eqref{first}.

The remaining terms in $dt_K$ can be formally written as
\be
\f{1+A_2z^2+A_4z^4}{1+B_2z^2}= 1+C_2z^2\ ,
\la{Form}
\ee
where $A_2,A_4$ are given in \eqref{A} and
\be
B_2=-\f{r'^2l^2}{4h^2r^2}
\la{Rest}.
\ee
The right hand side is determined by the identification of \eqref{B} with \eqref{K}:
\be
C_2\equiv -1\times \left[\f{dt_B}{dt_K}\ \f{Ha_0}{r'}\right]\ .
\la{C}
\ee
One has thus
to verify the identities
\be
B_2+C_2=A_2\quad\quad B_2C_2=A_4\ .
\la{Chek}
\ee
The first one leads to verifying
\be
Ha_0dt_B\equiv \dot{a_0}dt_B= r'dt_K l^2 \left[\f{r''}{2h^2r}-\f{h'r'}{2h^3r}-\f{r'^2}{4h^2r^2}\right]\ .
\la{second}
\ee
The equality \eqref{second} comes from the identification of both terms as
exact derivatives namely
\be
\dot{a_0}dt_B\equiv da_0 = r'dt_K \left[\f{r''}{2h^2r}-\f{h'r'}{2h^3r}-\f{r'^2}{4h^2r^2}\right]l^2 \equiv\left(\f{r'^2}{4h^2r}
\right)'l^2 dt_K\equiv d\left(\f{r'^2l^2 }{4h^2r}
\right)\ ,
\la{be}
\ee
which comes from \eqref{Id} giving $a_0=\f{r'^2l^2 }{4h^2r}$\ .

The last equation reads
\be
\f{r'^2}{4h^2r^2}\ dt_B= hdt_K \left[\f{r'^2r''}{8h^4r^3}-\f{r'^4}{16h^4r^4}-\f{h'r'^3}{8h^5r^3}\right]\ ,
\la{last}
\ee
which leads  to \eqref{second} and is thus verified.

Note that the metric shows a spatial singularity at
\be
z= \left\vert \f {2hr}{r'}\right\vert =\left\vert \frac{Hl}{2}\right\vert  ,
\la{Sing}
\ee
which is inherent to the choice $\Ccal=0=1/{z_0},$ as already noticed in footnote 1. In particular, in the pure de Sitter case with $\dot H=0$, the singularity is a conical singularity where both $ a(z,t_B)$ and $n(z,t_B)$ vanish simultaneously.

Finally  one finds, using \eqref{first}
\be
h\f {dt_K}{dt_B}=\epsilon_K \f 12\left(\frac{H^2l^2}{2}+\dot{H}l^2\right)
\la{tK}
\ee
which gives back \eqref{tau} upon identifying the holographic proper time as $d\tau= \vert h\vert dt_K$.
It will be  useful for the holographic correspondence to remark that Eq.\eqref{tK}  allows one to determine the sign of
$\epsilon_K$ in the definition \eqref{Id}.
Indeed, using \eqref{DotH}, one gets
\be
h\f {dt_K}{dt_B}=-\epsilon_K \frac {H^2l^2}4 (2+3\omega)\ .
\la{tK1}
\ee
Hence, one finds ${\rm sign}(h)=\epsilon_K$ for $\omega < -2/3$ and ${\rm sign} (h)=-\epsilon_K$ ${\rm for}\ \omega > -2/3.$

\subsection{Cosmological holography}
We are now interested in the more general case where $\Ccal\ne 0$. This modifies the metric, see \eqref{Eqb}. In particular, we have
\begin{equation}
a^2=\f {a_0^2}{z^2} \left( -z^2\frac{H^2l^2}{2} + \frac{1}{2} (z^4 -{1})\sqrt{1-\frac{\Ccal}{a_0^4}+H^2l^2} +\frac{1}{2}(z^4 +{1}) (1+\frac{H^2l^2}{2})\right)
\end{equation}
to be identified with
\begin{equation}
a^2= \frac{r^2}{z^2} \left(\left[1-\frac{l^2 r'^2 z^2}{4h^2r^2}\right]^2 + \frac{z^4}{4r^4 z_0^4}\right)
\end{equation}
from which we deduce that
\begin{equation}
h=\epsilon_K \frac{r'}{Ha_0}
\la{h}\end{equation}
together with
\begin{equation}
r^2=\frac{a_0^2}{2}\left(1-\sqrt{1-\frac{\Ccal}{a_0^4}+H^2l^2}+\frac{H^2l^2}{2}\right)
\end{equation}
and finally we link the two remaining parameters on each side of the mapping, namely
\begin{equation}
\Ccal\equiv\frac{1}{z_0^4}\ .
\end{equation}
This general holographic mapping is valid for all $Hl$ and ${\cal C}$, i.e. going beyond the low energy with no dark radiation regime studied in the previous section. We have derived it assuming that the dynamics on the physical brane are known. On the other hand, if the dynamics on the holographic brane are specified by $h(t_K)$ and $r(t_K)$, the physical dynamics can be reconstructed using
\begin{equation}
\dot a_0 = \epsilon_K \frac{dr}{ hdt_K}
\end{equation}
and
\begin{equation}
a_0^2= r^2 -\frac{1}{2} (\frac{r'l}{h})^2 + \frac{{\cal C} - (\frac{r' l}{h})^2}{4(r^2 -\frac{1}{2} (\frac{r'l}{h})^2)}
\end{equation}
Using the Friedmann equation on the physical brane, one can then reconstruct the energy density $\rho$. The sign of $\epsilon_K$ is intimately linked to the reconstructed equation of state and the sign of $h$ as we will see shortly, implying that the duality is a one to one correspondence between the quantum properties of the holographic brane and the classical features of the physical brane.

In general we find that the holographic energy density is
\begin{equation}
\rho_H= \frac{3N_c^2}{8\pi^2l^4} \frac{a_0^4}{r^4} \left( \frac{\Ccal}{a_0^4} + \frac{H^4l^4}{4}\right)
\end{equation}
and the isotropic pressure is
\begin{equation}
p_H=\frac{\rho_H}{3 } -\epsilon_K\ \frac{N_c^2}{8\pi^2}\ \left(\frac{a_0}{r}\right)^3 H^2\frac{\ddot a_0}{a_0}\ \frac{da_0}{dr}
\label{anom}
\end{equation}
where, {{from Eq.\eqref{h}}}
\begin{equation}
\frac{hdt_K}{dt_B}\equiv \epsilon_K \frac{dr}{da_0}
\end{equation}
Notice the contribution of the dark radiation to the energy density and the presence of the acceleration on the physical brane in correspondence with the pressure on the holographic brane.
Using the exact Friedmann equation, we find that
\begin{equation}
\frac{r^2}{a_0^2}= \frac{1}{4}\left\{ \left(1-\frac{\rho_b}{\rho_\Lambda}\right)^2 +\frac{\Ccal}{a_0^4}\right\}\ .
\end{equation}
Upon using  $\rho_b=\rho_\Lambda +\rho$ we obtain
\begin{equation}
\frac{r^2}{a_0^2}= \frac{1}{4}\left\{ \frac{\rho^2}{\rho_\Lambda^2} +\frac{\Ccal}{a_0^4}\right\}
\label{r}
\end{equation}
showing both the dependence on matter and dark radiation. This is the most convenient form of the holographic mapping. In particular, one can distinguish two regimes at low energy. The first one corresponds to the case where the dark radiation term is negligible. In this case $r\propto a_0^{-2-3w}$ and therefore we retrieve the fact that $\epsilon_K =-{\rm sign}(h)$ when $2+3w>0$. In the other case where dark radiation is relevant, we find that $r\propto a_0^{-1}$ and therefore $\epsilon_K=-{\rm sign}(h)$ too. In the following we will focus on the regimes where
\begin{equation}
r\sim a_0^{-\beta}
\end{equation}
where $\beta=1$ or $\beta=2+3w$ depending on the dominant contribution in the holographic mapping. When applying the holographic map in the long time asymptotic regime where $a_0$ becomes large, we can distinguish two different behaviours:
\begin{eqnarray}
\beta&=& 2+3w, \quad w\le -\frac{1}{3}\nonumber \\
\beta&=&1,\quad w>-\frac{1}{3} \nonumber \\
\end{eqnarray}
Hence, in this case, one can relate the value exponent $\beta$ to the existence of an acceleration phase. When $w>-1/3$ in a decelerated phase, $\beta=-1$ whereas it is $w$-dependent in an accelerated phase. Notice that
\begin{equation}
\vert\beta\vert \approx 1
\end{equation}
in the cosmological eras with $w=0, 1/3$ or even in a accelerated phase with $w\approx -1$.

At low energy,  the role of dark radiation from the holographic point of view is crucial. Indeed, it is possible for dark radiation to be negligible in the energy budget of the universe, i.e. its contribution to $H^2l^2$ in the Friedmann equation is subdominant compared to the matter component $\frac{\cal C}{a_0^4} \ll \frac{\rho}{\rho_\Lambda}$, whereas its presence could dominate in the mapping between the cosmological expansion rate $a_0$ and the holographic one $r$. This is realised when $\frac{\cal C}{a_0^4}\gg \frac{\rho^2}{\rho_\Lambda^2}$. In particular, it could well be that the present acceleration phase of the universe corresponds to such a situation. Hence holographic properties are sensitive to the details of the bulk metric, i.e. the absence or presence of a term ${\cal C}$, in a way which is much more acute than the cosmological properties on the
physical brane. Note that in \cite{Kajan}, ${\cal C}$ corresponds to a black hole mass.

An important consequence of \eqref{anom}  is the intimate relation between the conformal anomaly on the gauge theory side and acceleration on the cosmological side:
\begin{equation}
{\rho_H}-{3 }p_H=  \frac{\ddot a_0}{a_0}\ \frac{3N_c^2}{8\pi^2}\ H^2\left(\frac{a_0}{r}\right)^3\  \frac{dt_B}{h dt_K}\ .
\label{anom1}
\end{equation}
which shows that the  sign of $h$ and the one of the acceleration $\ddot a_0$ are correlated with the one of ${\rho_H}-{3 }p_H.$  Therefore  the positivity of the left-hand side of \eqref{anom1}, which corresponds  to  the gauge theory having a
positive conformal anomaly,  i.e.  the bulk viscosity of its plasma fluid being positive {{(see the discussion in \cite{Kajan} in our holographic context)}}, implies  that
\begin{equation}
{\rm sign}(\ddot a_0)= {\rm sign}(h)
\end{equation}
Hence we find that the change of regime on the physical brane from a decelerated phase to an accelerated one can only occur when the lapse function on the holographic brane changes sign. Using the holographic map, we have seen that the sign of a $h$ is linked to the sign of $\epsilon_K$ and the equation of state $w$. Let us concentrate  on the main eras of the universe, i.e. $w=0, 1/3$ and $w\approx -1$. We find that in the decelerated eras $w=0,1/3$, the sign of $h$ must be negative and therefore $\epsilon_K=1$ implying that the holographic brane undergoes a contraction phase with $\frac{dr}{da_0} <0$. On the other hand when $w\approx -1$, we have $\epsilon_K=1$ too while $\frac{dr}{da_0}>0$. Therefore the holographic brane experiences a bounce from a contraction to an expansion phase. Moreover the change of sign of $\ddot a_0$ occurs exactly when $h=0$ or equivalently when $\frac{dr}{da_0}=0$.

More generally we can distinguish the different types of physical cosmology depending on the equation of state $w$. First of all, one can formulate
\eqref{r} as
\begin{equation}
\frac{r^2}{a_0^2}= \frac{1}{4}\left ( \frac{\rho_0^2}{\rho_\Lambda^2}\frac{1}{a_0^{6(1+w)}} +\frac{\Ccal}{a_0^4}\right )
\end{equation}
where $\rho=\frac{\rho_0}{a_0^{3(1+w)}}$. Considering the asymptotically large behaviour of the universe when $a_0$ goes to infinity, we find that the behaviour of $r$ is determined by the equation of state $w$. When $w<-1/3$ which corresponds to an accelerating universe, the term in $\rho^2/\rho_\Lambda^2$ dominates whereas the $\frac{\cal C}{a_0^4}$ term dominates for a decelerating universe. As a result, we find that $\epsilon_K=1$ for
$-1\le w\le -2/3$. Similarly, we have that $\epsilon_K=1$ for $w>-1/3$ while $\epsilon_K=-1$ when $w\in [-2/3,-1/3]$. Dynamically, this corresponds to
$\frac {dr}{da_0} >0$ when $w<-2/3$ and $\frac{dr}{da_0}<0$ for $w>-2/3$. Table \ref{sign} summarises this discussion.

\TABLE{
\la{sign}\caption[]{{\it ``Table of signs''.}  We summarise the determination of signs as a function of the value of $\omega$ (the equation of state on the cosmological brane). 1$^{rst}$ line: relative dominance of the matter density over the dark radiation at large $a_0$ in the holographic map;  2$^{nd}$ line: sign of the lapse function $h$; 3$^{rd}$ line: sign of the  cosmic acceleration;  4$^{th}$ line: expansion (+) $vs.$ contraction (-) of the dual plasma phase; 5$^{th}$ line: value of $\epsilon_K$ in \eqref{Id} and following.}
\begin{tabular}{|c|c|c|c|c|c|c|c|c|c|}
\hline $\omega$ & $-1$ & $ $ & $-\f 23$ & $ $ & $-\f 13$ & $ $& $0$ & $ $& $\f 13$ \\
\hline $\rho\ {\rm  vs.}\ {\cal C}$ & $ $ & $\rho$ & $ $ & $\rho$& $ $ & ${\cal C}$ & $ $ & ${\cal C}$ &\\
\hline ${\rm sign}(h)$ & $ $ & $+$ & $ $ & $+$& $ $ & $-$ & $ $ & $-$ &\\
\hline ${\rm sign}(\ddot a_0)$ & $ $ & $+$ & $ $ & $+$& $ $ & $-$ & $ $ & $-$ &\\
\hline ${\rm sign}(\f{dr}{da_0})$ & $ $ & $+$ & $ $ & $-$& $ $ & $-$ & $ $ & $-$& \\
\hline $\epsilon_K$ & $ $ & $+$ & $ $ & $-$& $ $ & $+$ & $ $ & $+$& \\
\hline
\end{tabular}
}

\subsection{Brane-to-brane duality}

We have thus made explicit a mapping between the analytic continuation near $z=0$ of the bulk metric of the brane cosmology model of \cite{Bine} with the solution of the 5d Einstein equations corresponding to the gravity dual of the $\N=4$ SYM gauge theory with spherically expanding matter. Within this correspondence, the 4d physical  cosmology defined on the brane at $y=0 \ (z=1)$ is  in one-to-one correspondence with the properties of an holographic  $\N=4$ SYM plasma spherically expanding in a curved 4-dimensional space. This is the result of the analytic continuation of the bulk metric initially defined in  $y\ge 0$ ($z\ge 1$) to its $ Z_2 $ symmetric with $y\le 0$ ($0\le z\le 1$), see \eqref{Eqb}.

The metric \eqref{Eqb} allows  one to identify the physics on the holographic brane at $z=0$. Viewed on the unit interval $z\in \left [ 0,1\right ]$, the configuration is the following. The holographic brane at $z=0$ with its energy momentum tensor carries a spherically symmetric and expanding configuration of the $\N=4$ supersymmetric gauge theory, this theory being dual to the bulk theory with the aforementioned metric. Now this metric itself is such that the boundary brane located at $z=1$ carries a cosmological background characterized by a boundary energy momentum tensor. In this sense, we have realised a {\it brane-to-brane} duality through the holographic properties of  the bulk metric, $cf.$
Fig.\ref{1}.

This duality can be reinforced by noticing the following  form of \eqref{anom}. Using  the relations $dr/da_0= \epsilon_K hdt_K/dt_B$ and  $\sqrt{-g_B}= a^3,\  \sqrt {-g_K}=\vert h\vert r^3,$
 \eqref{anom} is equivalent to
\begin{equation}
\sqrt{-g_B}\, dt_B\  H^2 \frac{\vert \ddot a_0\vert }{a_0} = \sqrt{-g_K}\, dt_K\ \frac{8\pi^2}{3N_c^2}\ \left(\rho_H- 3 p_H\right)\ ,
\la{anom2}\end{equation}
Hence the acceleration of the physical brane is related to the deviation from conformality on the holographic brane. It is worth noting that
upon integrating \eqref{anom2}, and assuming convergence as we will discuss later on, one can  write
\begin{equation}
\int d^3 xdt_B \sqrt{-g_B}\ \vert \ddot a_0 \vert H^2 = \int d^3 x\,  dt_K \sqrt{-g_K}\ \frac{8\pi^2}{3N_c^2} (\rho_H-3p_H)\ .
\la{anom3}
\end{equation}

The conformal anomaly being an extensive quantity and $N_c^2$ being proportional to the number of degrees of freedom on the gauge theory side, in the large $N_c$ limit, let us define the  reduced anomaly
\begin{equation}
{\cal A}\equiv \lim_{N_c\to \infty} \frac{8\pi^2}{3} \frac{ \rho_H-3p_H}{N_c^2}= {\cal E}^4,
\la{Anom}
\end{equation}
where we introduced the fourth root ${\cal E}\equiv {{\cal A}}^{1/4}$ which has  dimension-one in energy units. Therefore $\cal E$ is  finite in the limit of large $N_c$ implying that
\begin{equation}
dt_B \sqrt{-g_B}\ H^2 \frac{\vert \ddot a_0\vert }{a_0}= dt_K \sqrt{-g_K}\  {\cal E}^4\ .
\la{Fini}\end{equation}

Eq.\eqref{Fini} has an interesting physical interpretation. The left hand part of the equality, corresponding to the gravity side of the Gauge/Cosmology duality is proportional to the acceleration of the universe which is the key experimental observable in the case of  dark energy. The right-hand part of \eqref{Fini}, corresponding to the microscopic side of the duality, is physically related to the bulk viscosity of the relativistic gauge fluid $via$ the conformal anomaly.

\subsection{Conformal anomaly $vs.$ e-foldings}

The geometrical  properties of the metric \eqref{Eqb} can be discussed referring to the results of \cite{Kajan}. In this study, the holographic correspondence of a gauge fluid isotropically expanding with  time in a  $1\!+\!3$ dimensional space with a time-dependent $AdS_5$ black hole geometry has been spelt out. This was motivated by an  extension of the similar structure found at large proper-time \cite{Jani} for the  $1\!+\!1$-dimensional  flow of a $\N=4$ SYM fluid.

Here, it can be seen that the metric element $a(z,t_B)$ does not vanish anymore when ${\cal C}\ne 0$.
We are interested in finding horizons in the bulk
which would  appear when ${da}/{da_0}=n=0$ or equivalently when ${da^2}/{da_0}=0$, this can be  realised for two values in the bulk
\begin{equation}
z_\pm^2= \frac{\frac{\ddot a_0}{a_0}l^2 \pm \sqrt{\Delta}}{ 2+ \frac{\ddot a_0}{a_0}l^2 +  \frac{ 2-H^2l^2 -l^2 \frac{\ddot a_0}{a_0}}{\sqrt{1-\frac{\Ccal}{a_0^4}+ H^2l^2}}}
\end{equation}
where
\begin{equation}
\Delta= -4 - 4\frac{\ddot a_0}{a_0}l^2 + \frac{ \left(2-H^2l^2 -l^2 \frac{\ddot a_0}{a_0}\right)^2}{1-\frac{\Ccal}{a_0^4}+ H^2l^2}
\end{equation}

These are potentially moving horizons, as in \cite{Kajan}. However, in the low energy cosmological set-up we are considering, it is possible to show that  the discriminant $\Delta$ is negative implying that there is no horizon in between the holographic and the physical brane. Hence the metric relation between the holographic and the physical brane is everywhere smooth. Indeed, using the physical requirements $\frac{\Ccal}{a_0^4}\ll H^2l^2 \ll 1$ valid at low energy, one finds
\begin{equation}
\Delta \sim 4\left(C/a_0^4 -2H^2l^2 -2l^2 a_0''/a_0\right) \sim -3(1+w)H^2l^2
\la{Hori}
\end{equation}
which stays negative provided $w>-1$, the cases of interest here. Note that, in  contrast with the cases discussed in \cite{Kajan}, the dark radiation term does not induce the existence of a black hole in the 5-dimensional bulk. However, the black hole structure analytically continued in the complex $z$ plane is essential for the holographic correspondence.

Let us now focus on the link between the  acceleration of the universe  and the conformal anomaly.
Upon using the equations of state,   we find
\begin{equation}
  {\cal E}d\tau  = \left\vert \beta^3\ \frac{(1+3w)}{2}\right\vert ^{1/4} H dt_B
\end{equation}
in an expanding universe where $H>0$ and
 the holographic proper time is $d\tau=\vert h\vert  dt_K$ leading to
\begin{equation}
  {\cal E}d\tau \equiv   \left\vert \beta^3\ \frac{ \left \vert 1+3w \right \vert }{2}\right\vert ^{1/4} {dN_B}
\la{Rela}
\end{equation}
where the number of e-folds on the physical brane is such that $dN_B=d\ln a_0$.

The relation \eqref{Rela} could acquire  a nice physical interpretation. Indeed, let us consider  time-dependent processes on both sides of the correspondence. On the gauge theory side one may consider a time interval
 with a positive non zero conformal anomaly, such as a phase transition. We immediately find that, by integrating both sides over the duration of the gauge theory process, the left hand side is given by $\Delta \tau  \Delta {\cal E}$ where the typical reduced conformal anomaly is defined via
\begin{equation}
\int_{\tau_{\rm ini}}^{\tau_{\rm ini}+ \Delta \tau}\!\! \!{\cal E}\, d\tau\  \equiv \Delta {\cal E} \Delta \tau  \ .
\end{equation}
Similarly, the right hand side of the duality equation \eqref{Rela} becomes
\begin{equation}
\int_{N_{\rm ini}}^{N_{\rm ini}+ \Delta N} \left\vert \frac{1+3w}{2}\right\vert ^{1/4} {dN_B}\equiv  \left \vert \frac{1+3w_{\rm av}}{2}\right\vert^{1/4}  \Delta N
\end{equation}
defining the averaged equation of state  $ w_{\rm av}$ on the physical brane. We have used the fact that $\beta\approx 1$ for all the relevant cosmological eras. Collecting all this, we find
\begin{equation}
\Delta \tau  \Delta {\cal E}=  \left\vert \frac{1+3w_{\rm av}}{2}\right\vert^{1/4}  \Delta N
\end{equation}
This equation is a quantum-classical duality. It relates the quantum phenomena on the gauge theory side, involving processes
with a finite duration $\Delta \tau$  and a typical energy scale  $\Delta {\cal E}$ to a purely classical quantity,
i.e. the number of e-foldings on the physical brane. It is valid for any equation of state $w_{\rm av}$, hence we find that the duration of the cosmological eras dual to microscopic phenomena on the holographic brane are related to a Heisenberg uncertainty relation in the dual gauge theory.

This relation is even more striking as for a typical quantum gauge process we have that
$\Delta \tau =O(\Lambda_G^{-1})$ where $\Lambda_G$ is the strong interaction scale of the non-conformal gauge theory and $\Delta {\cal E} =O(\Lambda_G)$ is the typical energy  released by  any quantum gauge process. As a result we find the quantum gauge processes saturate a Heisenberg uncertainty relation and
\begin{equation}
\Delta \tau  \Delta {\cal E}\sim \Delta N = O(1)
\la{Qcla}
\end{equation}
where we have used $w_{\rm av}\sim -1$ when the phase transition leads to an accelerated phase. It immediately implies that the duration of the accelerated expansion phase on the physical brane due to the
non-conformal transition on the holographic brane is of the order of a few e-foldings.
Moreover the quantum Heisenberg-type uncertainty relation on the {\it holographic} brane is  in one-to-one correspondence with a classical  gravity property on the {\it cosmological} brane.

As an illustration, let us consider the case of a transition of the QCD-type on the holographic brane. A transition of this kind on the holographic brane may have happened at the onset of the acceleration era of the universe at a redshift of a few, i.e. when the universe was around 10 billion years younger. As a result, the length of time since the beginning of the acceleration phase is of the order of the age of the universe.
The observed fact that $\Delta N=O(1)$ during the acceleration phase   has to be compared with the  microscopic QCD-like features of the transition whereby the typical energy  scale is of the order $\Lambda_{QCD}\approx 200\ MeV $ and a typical interaction time is of order $\Delta \tau \approx 10^{-23} \ {\rm second}\ \sim \ 5 \ GeV^{-1}\! \approx \Lambda^{-1}_{QCD}.$ Hence a possible interpretation of the recent accelerated phase of the universe could be to argue that, at its onset, there could be a phase transition on the holographic brane of the QCD type. Of course, this would require that the dynamics on the holographic brane were accompanied with a passage from a contracting phase to an expansion era.  More work along these lines is certainly worth pursuing.

Let us further speculate. If the QCD-like process at the origin of this dynamical process on the holographic brane were to be interpreted as the result of a multi-particle collision with a short compression phase  followed by a long relaxing phase and a deviation from conformality, the resulting dual cosmological configuration on the cosmological brane would be that of an accelerated universe whose duration would simply be a few efoldings.

\section{Conclusion and outlook}
\la{Fine}
Let us summarise briefly our approach:
\bi
\ii
We have shown using an explicit example, that brane-world cosmology with a single brane in 5d has a holographic dual in terms of a
$\N=4$ SYM gauge field theory on a holographic brane at the boundary of the $Z_2$-symmetric mirror bulk, thanks to the AdS/CFT correspondence with the bulk metric.

\ii
 A brane-to-brane Gauge/Cosmology duality establishes a Quantum/Classical  scale-invariant relationship  between a Heisenberg-type uncertainty relation on the holographic brane and the  number of e-foldings of the dual cosmology on the physical brane.
\ei

A relation between the conformal anomaly of a CFT and the cosmological acceleration has been already proposed in Ref.\cite{Siop}. The major difference of our approach is the introduction of two distinct branes, the {\it cosmological} one being different from the {\it holographic} one. In \cite{Siop} they are identical, the cosmological set-up being related to appropriate boundary conditions on the brane.

For completeness, let us summarise how the brane-to-brane Gauge/Gravity duality, with two  branes {{{\it distinctly}}} separated in the mirror image of the bulk, see Fig.\ref{1}, allows one to obtain many interesting features: one gets i) the identification of the dual Gauge Theory to a 4d cosmological background as the $\N=4$ SYM in an expanding background, ii) the scale-invariance  of the duality relation which implies that the overall scale of any gauge theory process does not impinge on the energy scales of the cosmological phenomena.

The Quantum/Classical correspondence illustrated  in our approach by the duality relation \eqref{Qcla} seems general enough to investigate the possibility of its extension to more general Gauge/Gravity set-ups. As an outlook and a  qualitative illustration of the possible wider validity of the gauge/cosmology brane-to-brane duality, let us consider a conformal theory and a non-conformal process occurring on the gauge theory side leading to a time dependence ${\cal{E}}\ne 0 $ during a finite amount of  proper time. Assuming that ${\cal E}$ can be treated a small perturbation, we find that the equation of  state on the physical brane is temporarily non-vanishing with
\begin{equation}
\left\vert \frac{1+3w}{2}\right\vert ^{1/4}\!\!\!\!(\tau)=   {\cal E}(\tau) \frac{d\tau}{dN_B}. \end{equation}
The physical process violating conformality has a typical energy scale $E$ and a typical duration $E^{-1}$.
As before, the variation of the physical equation of state is independent of the energy scale $E$.
Hence, we  have exhibited a mechanism whereby the effect of quantum
phenomena of energy $E$ is not to generate a large  cosmological constant of order $E^4$. On the contrary, the energy scale is irrelevant. All that
matters is the dynamics of the quantum process, leading to a time variation of the equation of state going from no acceleration in the past to no acceleration in the future.

{{As a final comment, it would  be interesting to analyse the passage from the  {\it deceleration} eras to an {\it acceleration} phase of the universe. We have seen that, imposing that the bulk viscosity of the holographic fluid is always positive, deceleration  would correspond to a  negative lapse function $h$ and a negative $ \frac{dr}{da_0}<0$ whereas acceleration is associated to a positive lapse function and $\frac{dr}{da_0}>0$ when the universe is dominated by a cosmological constant. In particular, the transition from a deceleration to an acceleration phase is linked to a bounce in the holographic expansion rate and a change of sign of the lapse function. It is tempting to draw an analogy with the elliptic flow of strong interaction collisions. Indeed, if the holographic scale factor $r(t_K)$ is understood as describing the typical size of the interaction region, then the bounce on the holographic  brane can be seen as the compression followed by the relaxation phase of the collision. Of course, this problem deserves further study.

We hope that  further developments of the brane-to-brane duality approach may tighten the links between a microscopic theory and a fully realistic cosmological scenario. For instance, describing the inflationary era by comparing with the evolution of a gauge fluid would be interesting. Moreover a more quantitative application of the duality to the recent acceleration of the universe is certainly worth pursuing.

\section*{Acknowledgements}
We would like to warmly thank Romuald Janik for a fruitful collaboration on time-dependent holography and his participation in the starting stage of the present study. We would also like to thank D. Langlois and M. Saridakis for useful comments. One of us (Ph.~B.) would like to thank the EU Marie Curie Research \& Training network ``UniverseNet" (MRTN-CT-2006-035863) for support.

\end{document}